\documentclass[12pt,a4paper]{article}
	
\usepackage{a4wide}
\usepackage{amsmath}
\usepackage{amssymb}
\usepackage{amsfonts}
\usepackage{epsfig}
\usepackage{exscale}
\usepackage{float}
\usepackage{bbm}
\usepackage[numbers,sort&compress]{natbib}

\newcommand{\Z}{{\mathbb{Z}}}
\newcommand{\R}{{\mathbb{R}}}
\newcommand{\C}{{\mathbb{C}}}

\newcommand{\1}{{\mathbbm{1}}}
\newcommand{\0}{{0}}

\newcommand{\p}{\partial}

\makeatletter
\@addtoreset{equation}{section}
\makeatother

\title{\vskip-2cm Majorana Fermions in a Box}

\author{M.\ H.\ Al-Hashimi$^{a,b}$, A.\ M.\ Shalaby$^a$, and 
U.-J.\ Wiese$^{b}$
\footnote{Contact information: M.\ H.\ Al-Hashimi: hashimi@itp.unibe.ch,
+41 31 631 8878; A.\ Shalaby, amshalab@qu.edu.qa, +974 4403 4630; 
U.-J.\ Wiese, wiese@itp.unibe.ch, +41 31 613 8504.}
\\ \\
{\small $^a$ Department of Mathematics, Statistics, and Physics} \\
{\small Qatar University, Al Tarfa, Doha 2713, Qatar} \\
{\small $^b$ Albert Einstein Center for Fundamental Physics,
Institute for Theoretical Physics} \\
{\small Bern University, Sidlerstrasse 5, CH-3012 Bern, Switzerland}}

\begin{document} 

\maketitle

\vspace{-1cm}

\begin{abstract} \normalsize

Majorana fermion dynamics may arise at the edge of Kitaev wires or 
superconductors. Alternatively, it can be engineered by using trapped ions or
ultracold atoms in an optical lattice as quantum simulators. This motivates the 
theoretical study of Majorana fermions confined to a finite volume, whose 
boundary conditions are characterized by self-adjoint extension parameters.
While the boundary conditions for Dirac fermions in $(1+1)$-d are characterized
by a 1-parameter family, $\lambda = - \lambda^*$, of self-adjoint extensions,
for Majorana fermions $\lambda$ is restricted to $\pm i$. Based on this result,
we compute the frequency spectrum of Majorana fermions confined to a 1-d 
interval. The boundary conditions for Dirac fermions confined to a 3-d region
of space are characterized by a 4-parameter family of self-adjoint
extensions, which is reduced to two distinct 1-parameter families for Majorana
fermions. We also consider the problems related to the quantum mechanical 
interpretation of the Majorana equation as a single-particle equation.
Furthermore, the equation is related to a relativistic Schr\"odinger equation 
that does not suffer from these problems.
 
\end{abstract}

\newpage

\section{Introduction}

Majorana fermions \cite{Maj37} result from Dirac fermions \cite{Dir28} by 
imposing a reality condition on the Dirac spinor \cite{Pal11}. As a result, 
Majorana fermions are neutral and are their own antiparticles. In the minimal 
version of the standard model of particle physics, neutrinos are electrically 
neutral left-handed Weyl fermions \cite{Wey29} charged under the electroweak 
$SU(2)_L \times U(1)_Y$ gauge symmetry. In this case, no renormalizable neutrino
mass terms exist, and thus, in this minimal theoretical 
framework, neutrinos are massless particles. Since the observation of neutrino
oscillations, it is known that neutrinos indeed must have a small non-zero mass.
When one extends the standard model by introducing additional right-handed 
neutrino fields, one can construct gauge invariant Dirac mass terms which 
involve the Higgs field and give rise to non-zero neutrino masses via the Higgs 
mechanism of electroweak symmetry breaking. Gauge invariance then requires that
the right-handed neutrino fields are neutral under all gauge interactions. This
in turn implies that one can also construct gauge invariant renormalizable
Majorana mass terms which do not involve the Higgs field and thus give rise to
neutrino masses, unrelated to the energy scale of electroweak symmetry breaking.
Since the right-handed component does not participate in the 
electroweak or strong gauge interactions, Majorana neutrinos are extremely 
weakly interacting. In particular, like any neutrino they easily penetrate 
even dense materials and can thus not be confined in any container. Still, in
some extensions of the standard model with extra spatial dimensions, neutrinos
may be confined to finite regions of the extra-dimensional space.

The confinement of Majorana neutrinos in finite regions of space is a more
important issue in condensed matter physics. In particular, Majorana fermions, 
which may emerge as edge modes of Kitaev wires \cite{Kit01} or of 
superconductors \cite{Ben13}, have been discussed in the context of topological 
quantum computation \cite{Kit03,Den02,Fre03,Fre04,Nay08,Bre08}. Majorana 
fermions may also arise in engineered systems, such as ultracold atoms in 
optical lattices or ion traps \cite{Cas11,Lam12,Mez13}. We take these systems 
as a motivation to investigate the Majorana equation, restricted to a finite 
region in space, using the theory of self-adjoint extensions \cite{Neu32,Ree75}.
In previous work, we have analyzed the Schr\"odinger, Pauli, and Dirac equations
in a similar manner \cite{AlH12,AlH15}. For example, the perfectly reflecting 
wall of a box that confines nonrelativistic Schr\"odinger particles without 
spin is characterized by a single self-adjoint extension parameter. The most 
general boundary condition for relativistic Dirac fermions (which generalizes 
the boundary conditions of the MIT bag model \cite{Cho74,Cho74a,Has78}) is 
characterized by a 4-parameter family of self-adjoint extension parameters
\cite{AlH12}. As we will show, imposing the 
Majorana reality condition on the corresponding Dirac spinor restricts the 
admissible values of the self-adjoint extension parameters. We then study the 
Majorana equation both in $(1+1)$ and in $(3+1)$ dimensions, with confining 
spatial boundary conditions.

The rest of this paper is organized as follows. In Section 2 we investigate the 
Majorana equation in $(1+1)$ dimensions, review its symmetries, and relate it to
a relativistic Schr\"odinger-type equation with a consistent quantum mechanical
single-particle interpretation. In Section 3 we study the self-adjoint extension
parameters that characterize a perfectly reflecting boundary. The Majorana 
equation is then solved for a particle confined to a finite interval. In Section
4 we extend these investigations to $(3+1)$ dimensions by reviewing the Majorana
equation and its symmetries, and by again constructing an equivalent 
relativistic Schr\"odinger-type equation. In Section 5 we construct a family of
self-adjoint extensions for $(3+1)$-d Majorana fermions, confined to a finite
region of space. Finally, Section 6 contains our conclusions.

\section{Majorana Fermions in $(1+1)$ Dimensions}

In this section we investigate the Majorana equation in $(1+1)$ dimensions. In
particular, we review its symmetry properties and investigate some problems
related to its quantum mechanical interpretation as a single-particle equation.

\subsection{The Majorana equation in $(1+1)$ dimensions}

Let us first consider the Dirac equation in $(1+1)$ dimensions
\begin{eqnarray}
&&i \partial_t \Psi(x,t) = (\alpha p c + \beta M c^2) \Psi(x,t), \quad
\Psi(x,t) = \left(\begin{array}{c} \psi_1(x,t) \\ \psi_2(x,t) \end{array}
\right), \nonumber \\
&&\alpha = \left(\begin{array}{cc} 0 & 1 \\ 1 & 0 \end{array}\right), 
\quad \beta = \left(\begin{array}{cc} 1 & 0 \\ 0 & -1 \end{array}\right). 
\end{eqnarray}
Here $M$ is the fermion mass, $c$ is the velocity of light, and we have put
$\hbar = 1$. A consistent choice of the $\gamma$-matrices is provided in the 
Dirac basis
\begin{equation}
\gamma^0 = \beta = \left(\begin{array}{cc} 1 & 0 \\ 0 & -1 \end{array}\right),
\quad \gamma^1 = \gamma^0 \alpha =
\left(\begin{array}{cc} 0 & 1 \\ -1 & 0 \end{array}\right),
\end{equation}
where the space-time metric is given by $g_{\mu\nu} = \mbox{diag}(1,-1)$.
Alternatively, we can use a Majorana basis
\begin{equation}
\widetilde \gamma^0 = 
\left(\begin{array}{cc} 0 & -i \\ i & 0 \end{array}\right),
\widetilde \gamma^1 = 
\left(\begin{array}{cc} i & 0 \\ 0 & -i \end{array}\right),
\end{equation}
in which the $\widetilde\gamma$-matrices have purely imaginary entries. The
Dirac and the Majorana basis are related by the unitary transformation
\begin{equation}
U = \frac{1}{\sqrt{2}} 
\left(\begin{array}{cc} 1 & -i \\ i & -1 \end{array}\right), \quad
\gamma^\mu = U \widetilde\gamma^\mu U^\dagger, \quad 
\Psi(x,t) = U \widetilde\Psi(x,t).
\end{equation}
In the Majorana basis, the Dirac equation is consistent with imposing the
reality condition $\widetilde\Psi(x,t)^* = \widetilde\Psi(x,t)$. In the Dirac
basis, the Majorana condition takes the form
\begin{eqnarray}
\label{Mcondition}
\Psi(x,t)&=&U \widetilde\Psi(x,t) = U \widetilde\Psi(x,t)^* =
U [U^\dagger \Psi(x,t)]^* = U U^T \Psi(x,t)^* \nonumber \\
&=&\frac{1}{2} \left(\begin{array}{cc} 1 & -i \\ i & -1 \end{array}\right)
\left(\begin{array}{cc} 1 & i \\ -i & -1 \end{array}\right) \Psi(x,t)^* =
\left(\begin{array}{cc} 0 & i \\ i & 0 \end{array}\right) 
\left(\begin{array}{c} \psi_1(x,t)^* \\ \psi_2(x,t)^* \end{array}\right) \
\Rightarrow \nonumber \\
\psi_1(x,t)&=&i \psi_2(x,t)^*, \quad \psi_2(x,t) = i \psi_1(x,t)^*.
\end{eqnarray}
Introducing $\psi(x,t) = \psi_1(x,t)$ the 2-component Dirac equation reduces to 
the 1-component Majorana equation
\begin{eqnarray}
\label{Majorana1d}
&&i \partial_t 
\left(\begin{array}{c} \psi(x,t) \\ i \psi(x,t)^* \end{array}\right) =
(\alpha p c + \beta M c^2)  
\left(\begin{array}{c} \psi(x,t) \\ i \psi(x,t)^* \end{array}\right) \
\Rightarrow \nonumber \\
&&i \partial_t \psi(x,t) =  M c^2 \psi(x,t) + c \partial_x \psi(x,t)^*.
\end{eqnarray}
Here we have used $p = - i \partial_x$. Unlike for the Schr\"odinger or Dirac 
equation, the right-hand side of the Majorana equation involves both $\psi(x,t)$
and $\psi(x,t)^*$. As a consequence, it can not be interpreted as an ordinary
quantum mechanical Hamiltonian acting on a wave function $\psi(x,t)$. In any 
case, a quantum mechanical single-particle interpretation is problematical 
already for the Dirac equation. Putting this caveat aside, one can still use 
the Dirac Hamiltonian as well as other quantum mechanical operators of the Dirac
theory, acting on constrained Majorana wave functions, to define expectation
values for Majorana fermions. For the expectation value of the energy one then
obtains
\begin{eqnarray}
\langle H \rangle&=&\int dx \left(\psi^*, - i \psi\right)
\left(\alpha p c + \beta M c^2\right)
\left(\begin{array}{c} \psi \\ i \psi^* \end{array}\right) \nonumber \\
&=&\int dx \left(\psi^*, - i \psi\right) 
\left(\begin{array}{cc} M c^2 & - i c \partial_x \\ - i c \partial_x & - M c^2
\end{array}\right) \left(\begin{array}{c} \psi \\ i \psi^* \end{array}\right) 
\nonumber \\
&=&\int dx \left(\psi^*, - i \psi\right) 
\left(\begin{array}{cc} M c^2 \psi + c \partial_x \psi^* \\ 
- i c \partial_x \psi - i M c^2 \psi^* \end{array}\right) \nonumber \\
&=&\int dx \left(\psi^* i \partial_t \psi + \psi i \partial_t \psi^*\right) = 
i \partial_t \int dx \ |\psi|^2 = 0. 
\end{eqnarray}
In the last step we have used the Majorana equation. As we will see in the next
subsection, the total ``probability'' $2 \int dx \ |\psi|^2$ is indeed 
conserved. As a consequence, the energy expectation value of a Majorana fermion 
state, evaluated with the Dirac Hamiltonian, always vanishes. The same is true 
for the momentum operator
\begin{eqnarray}
\langle p \rangle&=&\int dx \left(\psi^*, - i \psi\right)(- i \partial_x)
\left(\begin{array}{c} \psi \\ i \psi^* \end{array}\right) \nonumber \\
&=&\int dx \left(- i \psi^* \partial_x \psi - i \psi \partial_x \psi^*\right) = 
- i \int dx \ \partial_x |\psi|^2 = 0. 
\end{eqnarray}
Here we have used partial integration and we have assumed that the wave function
vanishes at spatial infinity. The expectation values of energy and momentum
vanish because a Majorana fermion is an equal weight superposition of positive 
and negative energy and momentum states. As a consequence, the solutions of the 
Majorana equation do not include stationary energy eigenstates with a unique 
(positive or negative) energy.

\subsection{Conserved ``probability'' current}

The Majorana equation is not invariant against multiplication of $\psi(x,t)$ by 
an arbitrary $U(1)$ phase, but only against a change of sign. As a result, 
fermion number is conserved only modulo 2. Interestingly, the Majorana equation 
still inherits the conserved current of the Dirac equation, 
\begin{eqnarray}
j^\mu(x,t)&=&\overline{\Psi}(x,t) \gamma^\mu \Psi(x,t) \ \Rightarrow \nonumber \\
\rho(x,t)&=&\overline{\Psi}(x,t) \gamma^0 \Psi(x,t) = 
\Psi(x,t)^\dagger \Psi(x,t) = |\psi_1(x,t)|^2 + |\psi_2(x,t)|^2, \nonumber \\
j(x,t)&=&c \overline{\Psi}(x,t) \gamma^1 \Psi(x,t) = 
c \Psi(x,t)^\dagger \gamma^0 \gamma^1 \Psi(x,t) = 
c \Psi(x,t)^\dagger \alpha \Psi(x,t) \nonumber \\
&=&c \left[\psi_1(x,t)^* \psi_2(x,t) + \psi_2(x,t)^* \psi_1(x,t)\right],
\end{eqnarray}
which, after imposing the Majorana condition eq.(\ref{Mcondition}), takes the 
form
\begin{equation}
\label{Majoranacurrent}
\rho(x,t) = 2 |\psi(x,t)|^2, \quad 
j(x,t) = i c \left[\psi(x,t)^{*2} - \psi(x,t)^2\right].
\end{equation}
Indeed, by using the Majorana equation (\ref{Majorana1d}), we obtain
\begin{eqnarray}
\partial_t \rho(x,t) + \partial_x j(x,t)&=&
2 \left[\psi(x,t)^* \partial_t \psi(x,t) + 
\psi(x,t) \partial_t \psi(x,t)^* \right] \nonumber \\
&+&2 i c \left[\psi(x,t)^* \partial_x \psi(x,t)^* - 
\psi(x,t) \partial_x \psi(x,t)\right] \nonumber \\
&=&- 2 i \psi(x,t)^* \left[M c^2 \psi(x,t) + c \partial_x \psi(x,t)^*\right] 
\nonumber \\
&+&2 i \psi(x,t) \left[M c^2 \psi(x,t)^* + c \partial_x \psi(x,t)\right] 
\nonumber \\
&+&2 i c \left[\psi(x,t)^* \partial_x \psi(x,t)^* - 
\psi(x,t) \partial_x \psi(x,t)\right] = 0.
\end{eqnarray}
Although, just like for the Dirac equation, a quantum mechanical single-particle
interpretation of the Majorana equation is problematical, and despite the fact
that Majorana fermion number is conserved only modulo 2, the continuity equation
implies that the total ``probability'' 
\begin{equation}
\int dx \ \rho(x,t) = 2 \int dx \ |\psi(x,t)|^2 = 1
\end{equation}
is conserved.

\subsection{Lorentz invariance}

Let us consider a Lorentz boost
\begin{eqnarray}
&&x' = \frac{x - v t}{\sqrt{1 - v^2/c^2}}, \quad 
c t' = \frac{c t - \frac{v}{c} x}{\sqrt{1 - v^2/c^2}} \ \Rightarrow \nonumber \\
&&\left(\begin{array}{c} c t' \\ x' \end{array}\right) = 
\gamma \left(\begin{array}{cc} 1 & - \beta \\ - \beta & 1 \end{array}\right)
\left(\begin{array}{c} c t \\ x \end{array}\right), \quad 
\beta = \frac{v}{c}, \quad \gamma = \frac{1}{\sqrt{1 - v^2/c^2}} = 
\cosh\theta \ \Rightarrow \nonumber \\
&&\left(\begin{array}{c} c t' \\ x' \end{array}\right) = \Lambda^{-1} 
\left(\begin{array}{c} c t \\ x \end{array}\right), \quad
\Lambda = \left(\begin{array}{cc} \cosh\theta & \sinh\theta \\ 
\sinh\theta & \cosh\theta \end{array}\right).
\end{eqnarray}
Under Lorentz boosts a Dirac spinor transforms as
\begin{equation}
\Psi'(x,t) = \left(\begin{array}{cc} \cosh\frac{\theta}{2} & 
\sinh\frac{\theta}{2} \\ 
\sinh\frac{\theta}{2} & \cosh\frac{\theta}{2} 
\end{array}\right) \Psi(x',t').
\end{equation}
For a Majorana spinor this implies
\begin{equation}
\label{LorentzMajorana}
\psi'(x,t) = \cosh\frac{\theta}{2} \ \psi(x',t') + 
i \sinh\frac{\theta}{2} \ \psi(x',t')^*.
\end{equation}
It is straightforward to show that the Majorana equation is indeed invariant
under this transformation.
 
\subsection{Parity, time-reversal, and charge conjugation}

Let us now consider the discrete symmetries P, T, and C for Majorana fermions in
one spatial dimension. For a Dirac fermion, the parity transformation P takes
the form
\begin{eqnarray}
&&^P\Psi(x,t) = \gamma^0 \Psi(-x,t) = \left(\begin{array}{cc} 1 & 0 \\ 0 & -1
\end{array}\right) \Psi(-x,t) \ \Rightarrow \nonumber \\
&&^P\psi_1(x,t) = \psi_1(-x,t), \quad ^P\psi_2(x,t) = - \psi_2(-x,t).
\end{eqnarray}
This is inconsistent with the Majorana condition 
$\psi_2(x,t) = i \psi_1(x,t)^*$. However, combining the Dirac parity operation
with a $U(1)$ phase multiplication by $i$ (which alone is not a symmetry of the
Majorana equation) we obtain the Majorana parity transformation
\begin{equation}
^P \psi(x,t) = i \psi(-x,t),
\end{equation}
which indeed leaves the Majorana equation invariant
\begin{eqnarray}
i \partial_t \ ^P\psi(x,t)&=&- \partial_t \psi(-x,t) = 
i M c^2 \psi(-x,t) + i c \partial_{-x} \psi(-x,t)^* \nonumber \\
&=&M c^2 i \psi(-x,t) + c \partial_x [i \psi(-x,t)]^* \nonumber \\ 
&=&M c^2 \ ^P\psi(x,t) + c \partial_x \ ^P \psi(x,t)^*.
\end{eqnarray}
As one would expect, under P the probability and current densities transform as
\begin{eqnarray}
^P\rho(x,t)&=&2 |^P \psi(x,t)|^2 = 2 |i \psi(-x,t)|^2 = \rho(-x,t), 
\nonumber \\
^P j(x,t)&=&i c \left[^P \psi(x,t)^{*2} - \ ^P \psi(x,t)^2\right] \nonumber \\
&=&i c \left[- \psi(-x,t)^{*2} + \psi(-x,t)^2\right] = - j(-x,t).
\end{eqnarray}

For a Majorana fermion, we define time-reversal as 
\begin{equation}
^T \psi(x,t) = \psi(x,-t)^*,
\end{equation}
which again leaves the Majorana equation invariant
\begin{eqnarray}
i \partial_t \ ^T\psi(x,t)&=&i \partial_t \psi(x,-t)^* = 
- i \partial_{-t} \psi(x,-t)^* \nonumber \\
&=&M c^2 \psi(x,-t)^* + c \partial_x \psi(x,-t) \nonumber \\
&=&M c^2 \ ^T\psi(x,t) + c \partial_x \ ^T \psi(x,t)^*.
\end{eqnarray}
Under time-reversal the probability and current densities transform as
\begin{eqnarray}
^T\rho(x,t)&=&2 |^T \psi(x,t)|^2 = 2 |\psi(x,-t)^*|^2 = \rho(x,-t), 
\nonumber \\
^T j(x,t)&=&i c \left[^T \psi(x,t)^{*2} - ^T \psi(x,t)^2\right] =
i c \left[\psi(x,-t)^2 - \psi(x,-t)^{*2}\right] \nonumber \\
&=&- j(x,-t).
\end{eqnarray}

Finally, let us consider charge conjugation C, which for a Dirac fermion takes
the form
\begin{equation}
^C \Psi(x,t) = \left(\begin{array}{cc} 0 & i \\ i & 0 \end{array}\right) 
\Psi(x,t)^* \ \Rightarrow \ ^C\psi_1(x,t) = i \psi_2(x,t)^*, \quad
^C\psi_2(x,t) = i \psi_1(x,t)^*. 
\end{equation}
As it should, this implies that a Majorana fermion is C-invariant
\begin{equation}
^C \psi(x,t) = \psi(x,t).
\end{equation}

\subsection{Propagation of wave packets}

By inserting the plane wave ansatz
\begin{equation}
\psi(x,t) = A \exp(i (k x - \omega t)) + B \exp(- i (k x - \omega t)),
\end{equation}
into the Majorana equation (\ref{Majorana1d}) one obtains
\begin{equation}
\omega = \sqrt{(M c^2)^2 + k^2 c^2}, \quad B = i A^* \frac{\omega - M c^2}{k c},
\end{equation}
such that the most general wave packet solution of the Majorana equation is
given by
\begin{equation}
\psi(x,t) = \int dk \left[A(k) \exp(i (k x - \omega t)) + 
i A(k)^* \frac{\omega - M c^2}{k c} \exp(- i (k x - \omega t))\right].
\end{equation}
The normalization condition, inherited from the Dirac equation, then takes the 
form
\begin{equation}
\langle\Psi|\Psi\rangle = \int dx \left(\psi^*, - i \psi\right) 
\left(\begin{array}{c} \psi \\ i \psi^* \end{array}\right) = 
2 \int dx \ |\psi|^2 = \frac{2}{\pi} \int dk \ |A(k)|^2 
\frac{\omega(\omega - M c^2)}{k^2 c^2}.
\end{equation}
We have seen that the expectation values of energy and momentum vanish because a
Majorana fermion is its own antiparticle. Let us now calculate the expectation 
value of the velocity operator
\begin{equation}
v = \partial_k \omega = \frac{k c^2}{\omega},
\end{equation}
which takes the form
\begin{equation}
\langle v \rangle(t) = \frac{2}{\pi} \int dk \ |A(k)|^2 
\frac{\omega - M c^2}{k} = \langle v \rangle(0),
\end{equation}
and hence is time-independent. It is straightforward but somewhat tedious to 
calculate the expectation value of the position operator and one obtains
\begin{eqnarray}
\label{xexpectation}
\langle x \rangle(t)&=&\langle x \rangle(0) + \langle v \rangle(0) t 
\nonumber \\
&+&\frac{1}{2 \pi} \Re \int dk \ A(-k) A(k) \frac{M c}{\omega k^2} 
(\omega - M c^2) \left[\exp(- 2 i \omega t) - 1\right] \nonumber \\
\langle x \rangle(0)&=&\frac{1}{2 \pi} \Re \int dk \ A(-k) A(k) 
\frac{M c}{\omega k^2} (\omega - M c^2)^2 \nonumber \\
&+&\frac{1}{\pi} \Im \int dk \ A(k) \partial_k A(k)^* 
\frac{\omega (\omega - M c^2)}{k^2 c^2}.
\end{eqnarray}
The oscillatory contribution to $\langle x \rangle(t)$ involving
$\exp(- 2 i \omega t)$ is reminiscent of ``Zitterbewegung''. This term is not
present for the propagation of wave packets following the nonrelativistic free
particle Schr\"odinger equation for which 
$\langle x \rangle(t) = \langle x \rangle(0) + \langle v \rangle(0) t$ 
\cite{Ehr27}.

\subsection{Relation of the Majorana equation to a relativistic \\
Schr\"odinger equation}

As we discussed before, it is well known that a quantum mechanical 
single-particle interpretation of the Dirac or Majorana equation is 
problematical. The right-hand side of the Majorana equation cannot even be 
viewed as a quantum mechanical Hamiltonian acting on a wave function, because 
it involves both $\psi$ and $\psi^*$. Let us map $\psi$ to a Schr\"odinger-type
wave function
\begin{equation}
\Phi(x,t) = \psi(x,t) + i \frac{\sqrt{(M c^2)^2 + p^2 c^2} - M c^2}{p c}
\psi(x,t)^*, \quad p = - i \partial_x,
\end{equation}
which obeys
\begin{eqnarray}
i \partial_t \Phi(x,t)&=&i \partial_t \psi(x,t) + 
i \frac{\sqrt{(M c^2)^2 + p^2 c^2} - M c^2}{p c} i \partial_t \psi(x,t)^*
\nonumber \\
&=&M c^2 \psi(x,t) + c \partial_x \psi(x,t)^* \nonumber \\
&-&i \frac{\sqrt{(M c^2)^2 + p^2 c^2} - M c^2}{p c}
\left[M c^2 \psi(x,t)^* + c \partial_x \psi(x,t)\right] \nonumber \\
&=&\sqrt{(M c^2)^2 + p^2 c^2} \left[\psi(x,t) +
i \frac{\sqrt{(M c^2)^2 + p^2 c^2} - M c^2}{p c} \psi(x,t)^*\right] \nonumber \\
&=&\sqrt{(M c^2)^2 + p^2 c^2} \ \Phi(x,t).
\end{eqnarray}
Remarkably, $\Phi$ obeys a relativistic Schr\"odinger equation with only 
positive energy states. In particular, the equation for $\Phi$ has a consistent
quantum mechanical single-particle interpretation, with 
$\sqrt{(M c^2)^2 + p^2 c^2}$ playing the role of the Hamiltonian. In the context
of point-particle relativistic quantum mechanics it is no problem that this
Hamiltonian is nonlocal (i.e.\ is contains derivatives of arbitrary order).

Interestingly, while the Majorana equation allows only a sign change of $\psi$,
the relativistic Schr\"odinger equation allows global $U(1)$ phase changes
\begin{equation}
\label{U1symmetry}
^\alpha\Phi(x,t) = \exp(i \alpha) \Phi(x,t),
\end{equation}
which give rise to a nonlocal conserved probability current that was
constructed in \cite{AlH14}. This current is not directly related to the 
conserved local Majorana current of eq.(\ref{Majoranacurrent}). One can invert 
the relation between $\psi$ and $\Phi$ to obtain
\begin{equation}
\psi(x,t) = \frac{1}{2 \sqrt{(M c^2)^2 + p^2 c^2}} \left[
\left(\sqrt{(M c^2)^2 + p^2 c^2} + M c^2\right) \Phi(x,t) + i p c \ \Phi(x,t)^*
\right].
\end{equation}
The simple $U(1)$ symmetry of eq.(\ref{U1symmetry}) then turns into the
complicated nonlocal transformation
\begin{eqnarray}
^\alpha\psi(x,t)&=&
\frac{1}{2 \sqrt{(M c^2)^2 + p^2 c^2}} 
\left[\left(\sqrt{(M c^2)^2 + p^2 c^2} + M c^2\right) \ ^\alpha\Phi(x,t) + 
i p c \ ^\alpha\Phi(x,t)^* \right] \nonumber \\
&=&\frac{1}{2 \sqrt{(M c^2)^2 + p^2 c^2}} 
\left[\left(\sqrt{(M c^2)^2 + p^2 c^2} + M c^2\right) \exp(i \alpha) \Phi(x,t) 
\right. \nonumber \\
&+&\left. i p c \exp(- i \alpha) \Phi(x,t)^* \right].
\end{eqnarray}
Similarly, the simple Lorentz transformation for a Majorana spinor of 
eq.(\ref{LorentzMajorana}) turns into a complicated nonlocal transformation
rule for $\Phi$, which is not very illuminating in the present context but may
be interesting to study in more details in the framework of relativistic quantum
mechanics of free particles (in contrast to quantum field theory) \cite{AlH09}.

The Schr\"odinger-type wave function $\Phi$ inherits its P and T symmetry 
properties from the Majorana ``wave function'' $\psi$
\begin{eqnarray}
^P \Phi(x,t)&=& \ ^P\psi(x,t) + i \frac{\sqrt{(M c^2)^2 + p^2 c^2} - M c^2}{p c} 
\ ^P\psi(x,t)^* \nonumber \\
&=&i \psi(-x,t) + \frac{\sqrt{(M c^2)^2 + p^2 c^2} - M c^2}{p c} \
\psi(-x,t)^* = i \Phi(-x,t), \nonumber \\
^T \Phi(x,t)&=& \ ^T\psi(x,t) + i \frac{\sqrt{(M c^2)^2 + p^2 c^2} - M c^2}{p c} 
\ ^T\psi(x,t)^* \nonumber \\
&=&\psi(x,-t)^* + i \frac{\sqrt{(M c^2)^2 + p^2 c^2} - M c^2}{p c} \
\psi(x,-t) = \Phi(x,-t)^*.
\end{eqnarray}

The introduction of $\Phi$ and its corresponding relativistic Schr\"odinger
equation may provide a consistent quantum mechanical single-particle 
interpretation of the Majorana equation. Based on this, one could evaluate new
expectation values. For example, when evaluated with $\Phi$ (rather than with
the Dirac spinor $\Psi$ that obeys the Majorana condition), one would obtain
$\langle x \rangle(t) = \langle x \rangle(0) + \langle v \rangle(0) t$ without
any additional contribution from ``Zitterbewegung'', such as the one in
eq.(\ref{xexpectation}). While this is interesting, it is not the subject
of the current paper. Here we stay with the original Majorana equation by
imposing the Majorana condition on a Dirac spinor, and accept the problems
of its quantum mechanical interpretation as a single-particle equation.

\section{Majorana Fermions Confined to an Interval}

In this section we investigate Majorana fermions in a 1-dimensional box. In
particular, we study the self-adjoint extension parameters that characterize a
perfectly reflecting boundary condition and we solve the Majorana equation for
a particle confined to an interval.

\subsection{Perfectly Reflecting Walls for Majorana Fermions}

It is well known to the experts, but only rarely emphasized in quantum mechanics
textbooks, that a quantum mechanical wave function need not necessarily vanish 
at a perfectly reflecting wall \cite{Bal70,Car90,Bon01,AlH12}. In fact, the 
most general perfectly reflecting Robin boundary condition is characterized by 
a self-adjoint extension parameter
$\gamma \in \R$ and takes the form $\gamma \Psi(0) + \partial_x \Psi(0) = 0$.
The standard textbook boundary condition $\Psi(0) = 0$ then corresponds to the
special case $\gamma = \infty$. The general Robin boundary condition ensures
that the nonrelativistic probability current vanishes at the boundary. This
implies that no probability is leaking out of the box. More than this is not 
required for a consistent unitary quantum mechanical evolution.

Let us begin by studying the $(1+1)$-d Dirac equation on the positive $x$-axis
with a perfectly reflecting boundary at $x = 0$ \cite{AlH12}. In order to 
investigate the Hermiticity of the Dirac Hamiltonian, we consider
\begin{eqnarray}
\langle \chi|H|\Psi\rangle&=&\int_0^\infty dx \ \chi(x)^\dagger 
\left[- c \alpha i \p_x + \beta m c^2\right] \Psi(x) \nonumber \\
&=&\int_0^\infty dx \ \left\{\left[- c \alpha i \p_x + \beta m c^2\right]
\chi(x)\right\}^\dagger \Psi(x) - i c \chi(0)^\dagger \alpha \Psi(0) \nonumber \\
&=&\langle \Psi|H|\chi\rangle^* - i c \chi(0)^\dagger \alpha \Psi(0),
\end{eqnarray}
which leads to the Hermiticity condition
\begin{equation}
\label{Dirachermiticity1d}
\chi(0)^\dagger \alpha \Psi(0) = 0.
\end{equation}
We now introduce the self-adjoint extension condition
\begin{equation}
\label{Diracselfadjointness1d}
\psi_2(0) = \lambda \psi_1(0), \quad \lambda \in \C,
\end{equation}
which reduces eq.(\ref{Dirachermiticity1d}) to
\begin{equation}
\chi(0)^\dagger 
\left(\begin{array}{cc} 0 & 1 \\ 1 & 0 \end{array}\right) \Psi(0) =
\left[\chi_1(0)^* \lambda + \chi_2(0)^*\right] \psi_1(0) = 0 \ 
\Rightarrow \ \chi_2(0) = - \lambda^* \chi_1(0).
\end{equation}
In order for $H$ to be self-adjoint, the domains of $H$ and $H^\dagger$ must 
coincide, i.e.\ $D(H) = D(H^\dagger)$. To achieve this, one must request
\begin{equation}
\lambda = - \lambda^*,
\end{equation}
i.e.\ $\lambda$ must be purely imaginary. Hence, for Dirac fermions in 1-d 
there is a 1-parameter family of self-adjoint extensions that characterizes a 
perfectly reflecting wall. The self-adjointness condition 
eq.(\ref{Diracselfadjointness1d}) implies
\begin{eqnarray}
j(0)&=&c \Psi(0)^\dagger \alpha \Psi(0) = c \Psi(0)^\dagger 
\left(\begin{array}{cc} 0 & 1 \\ 1 & 0 \end{array}\right) \Psi(0) = 
c \left[\psi_1(0)^* \psi_2(0) + \psi_2(0)^* \psi_1(0)\right] \nonumber \\
&=&c \left[\psi_1(0)^* \lambda \psi_1(0) + \psi_1(0)^* \lambda^* \psi_1(0)\right]
= 0.
\end{eqnarray}
Hence, as in the nonrelativistic case, the current $j(0)$ vanishes at the
perfectly reflecting wall.

Majorana fermions obey the additional constraint $\psi_2 = i \psi_1^*$, such 
that
\begin{equation}
\lambda \psi_1(0) = \psi_2(0) = i \psi_1(0)^* \ \Rightarrow \ |\lambda| = 1 \
\Rightarrow \ \lambda = \pm i.
\end{equation} 
Hence, Majorana fermions admit only two discrete self-adjoint extensions, no
longer a continuous 1-parameter family.

\subsection{Majorana fermion in a 1-d box}

Let us consider a 1-d box $x \in [-L/2,L/2]$ endowed with perfectly reflecting
boundary conditions. For Majorana fermions this means
\begin{equation}
\label{bc}
s_+ \psi(L/2) = \psi(L/2)^*, \quad s_- \psi(- L/2) = \psi(- L/2)^*, \quad
s_+, s_- = \pm 1. 
\end{equation}
In order to maintain parity symmetry, we demand that the parity transformed
field also obeys the boundary condition
\begin{eqnarray}
&&s_+ \ ^P\psi(L/2) = \ ^P\psi(L/2)^* \ \Rightarrow \ s_+ i \psi(- L/2) = 
[i \psi(- L/2)]^* = - i \psi(- L/2)^* \ \Rightarrow \nonumber \\
&&s_- = - s_+. 
\end{eqnarray}

We now make the ansatz
\begin{eqnarray}
\label{solution}
\psi(x,t)&=&A \exp(i (k x - \omega t)) + i A^* \frac{\omega - M c^2}{k c}
\exp(- i (k x - \omega t)) \nonumber \\
&+&B \exp(i (- k x - \omega t)) - i B^* \frac{\omega - M c^2}{k c}
\exp(- i (- k x - \omega t)),
\end{eqnarray}
with $\omega = \sqrt{(M c^2)^2 + k^2 c^2}$. Imposing the boundary conditions of
eq.(\ref{bc}) then implies
\begin{equation}
B = A \exp(i k L) \frac{\omega - M c^2 - i s_+ k c}{\omega - M c^2 + i s_+ k c},
\quad B = A \exp(- i k L) 
\frac{\omega - M c^2 - i s_- k c}{\omega - M c^2 + i s_- k c}.
\end{equation}
If we choose parity-violating boundary conditions with $s_- = s_+$, this implies
\begin{equation}
\exp(i k L) = \pm 1 \ \Rightarrow \ k = \frac{\pi}{L} n, \quad n \in \Z,
\end{equation}
which is equivalent to the nonrelativistic momentum quantization condition for 
the standard box boundary condition $\Psi(\pm L/2) = 0$. On the other hand,
using parity-symmetric boundary conditions with $s_- =- s_+$, one obtains the 
quantization condition
\begin{equation}
\exp(i k L) = \pm \frac{\omega - M c^2 + i s_+ k c}{\omega - M c^2 - i s_+ k c}
\ \Rightarrow \ \cos(k L) = \mp \frac{M c^2}{\omega}.
\end{equation}
Let us first consider the massless limit $M = 0$, $\omega = |k| c$, such that
\begin{equation}
\cos(k L) = 0 \ \Rightarrow \ k = \frac{\pi}{L}\left(n + \frac{1}{2}\right), \ 
n \in \Z.
\end{equation}
This solution also applies to massive fermions in the high-energy limit
$\omega \gg M c^2$. In the nonrelativistic limit, on the other hand, we obtain
\begin{equation}
\cos(k L) = \mp \frac{M c^2}{M c^2 + \frac{k^2}{2M}}.
\end{equation}
In the low-energy limit $\frac{k^2}{2 M} \ll M c^2$ this again leads to
\begin{equation}
\cos(k L) =\mp 1 \ \Rightarrow \ k = \frac{\pi}{L} n, \quad n \in \Z,
\end{equation}
It should be noted that the discrete $k$-values resulting from the quantization
conditions as well as the corresponding discrete frequencies
$\omega = \sqrt{(M c^2)^2 + k^2 c^2}$ do not yield stationary energy 
eigenstates. This is because the solution of eq.(\ref{solution}) is again a
superposition of states with positive and negative energy $\pm \omega$.

\begin{figure}[tbh]
\begin{center}
\epsfig{file=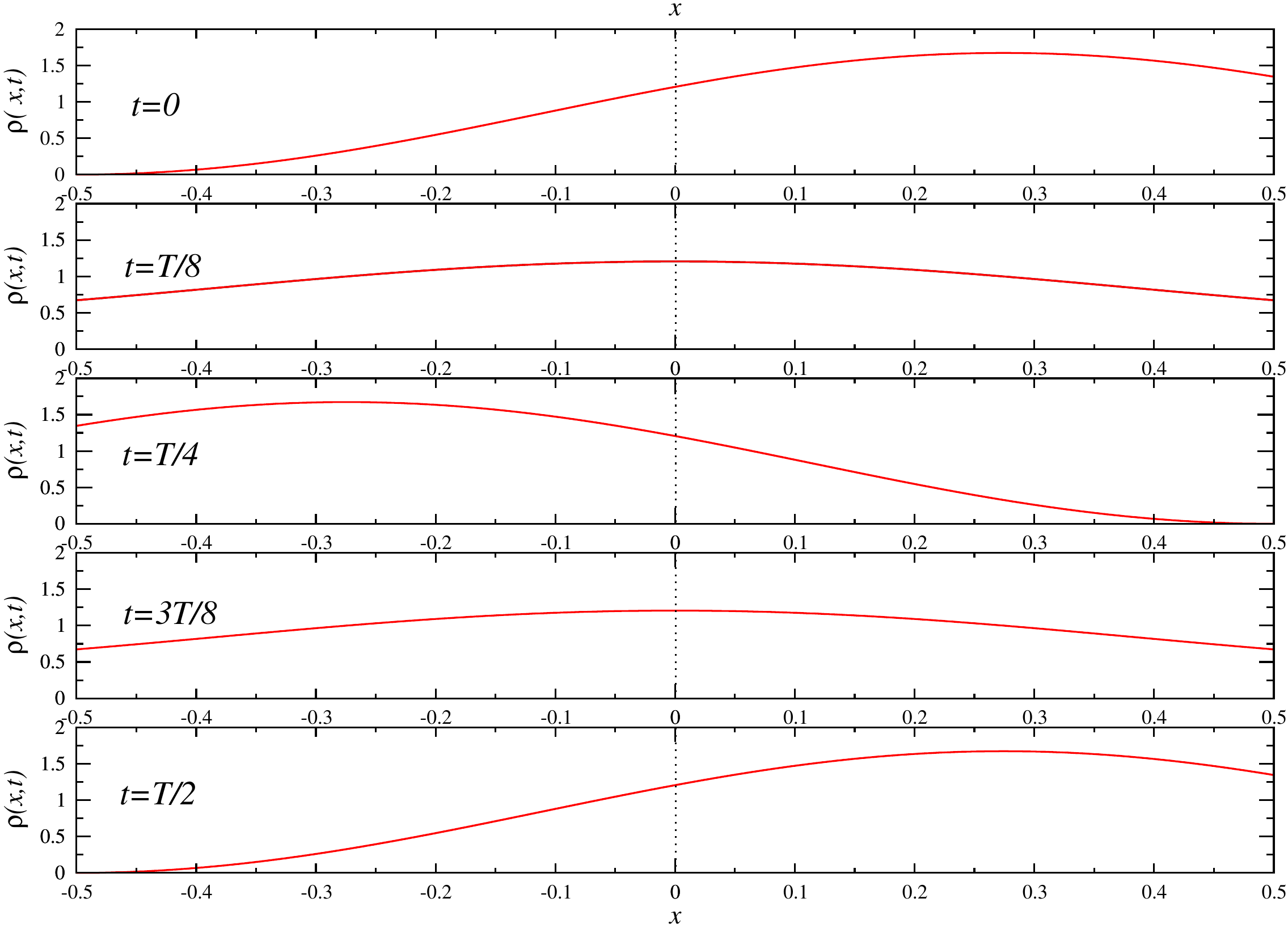,width=10cm}
\epsfig{file=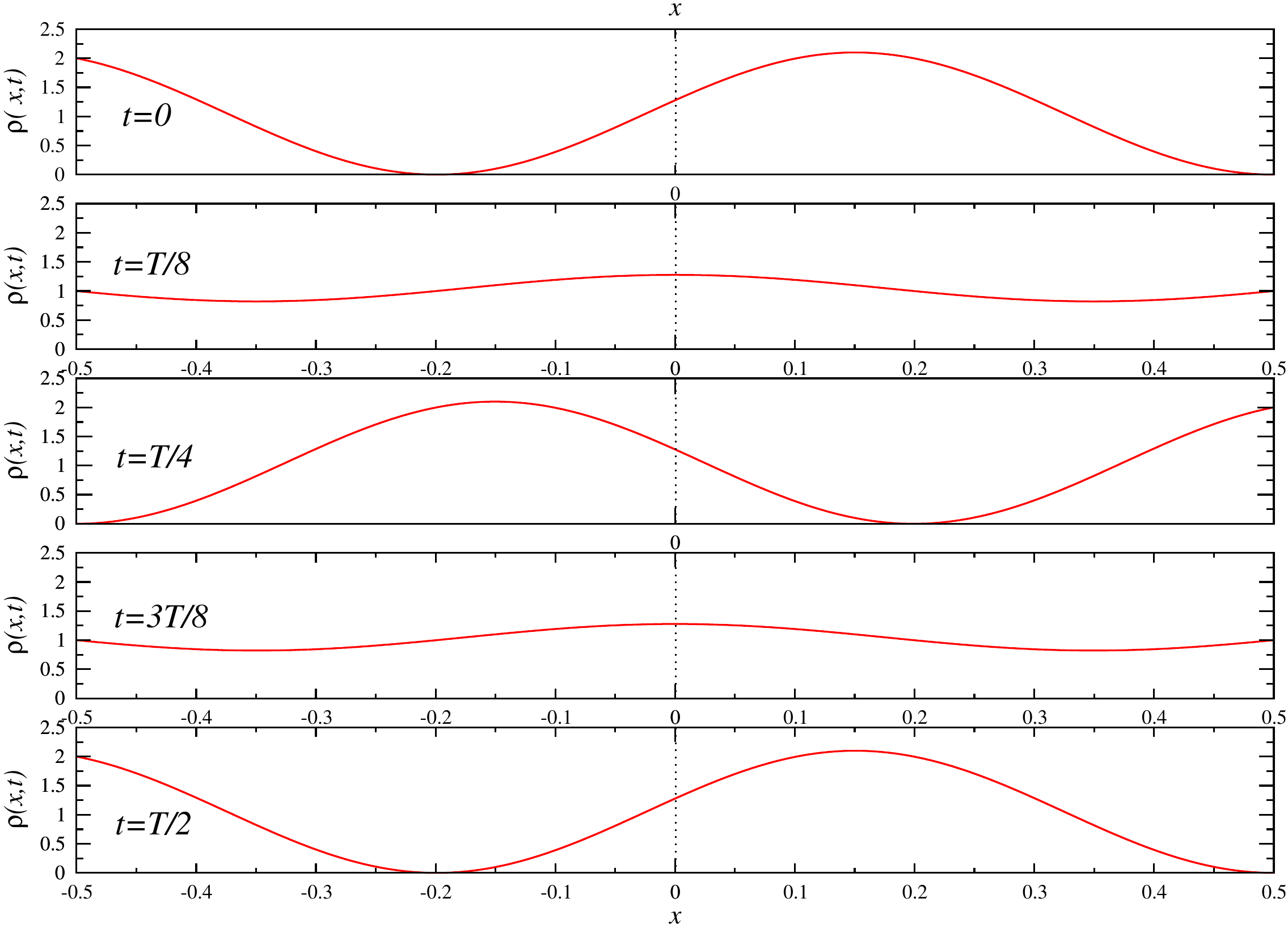,width=10cm}
\end{center}
\caption{\it Probability density $\rho(x,t)$ for a Majorana fermion confined to
a 1-d interval $x/L = [-\frac{1}{2},\frac{1}{2}]$ in the parity-respecting case
($s_+ = - s_-$) with $M L c = 1$ for various times $t$ in units of the period 
$T = 2 \pi/\omega$ of the wave function. Note that $\rho(x,t)$ is periodic in
time with period $T/2$. The probability density at $t = T/4$ is the parity
image of the initial density, i.e.\ $\rho(x,T/4) = \rho(-x,0)$. The state of 
lowest frequency is shown in the top panels, while the first excited state is 
shown at the bottom.}
\label{Prespecting}
\end{figure}

In the parity-respecting case ($s_+ = - s_-$), the probability density 
corresponding to the wave function of eq.(\ref{solution}) takes the form
\begin{eqnarray}
\label{Prespect}
\rho(x,t)&=&\frac{2 |A|^2 k}{c^2} \left[
k^2 c^2 + 2 k c (\omega - M c^2) \sin(2 k x) \cos(2 \omega t) \right. \nonumber
\\
&+&\left.
(k c + M c^2 - \omega)(k c - M c^2 + \omega) \cos(2 k x) + (M c^2 - \omega)^2
\right],
\end{eqnarray}
and the normalization factor is given by
\begin{equation}
\frac{1}{|A|^2} = 2 k c L \left[k^2 c^2 + (M c^2 - \omega)^2\right] + 
2 \sin(k L) (k c + M c^2 - \omega)(k c - M c^2 + \omega).
\end{equation}
The probability density of eq.(\ref{Prespect}) is illustrated in 
Fig.\ref{Prespecting}.

\begin{figure}[tbh]
\begin{center}
\epsfig{file=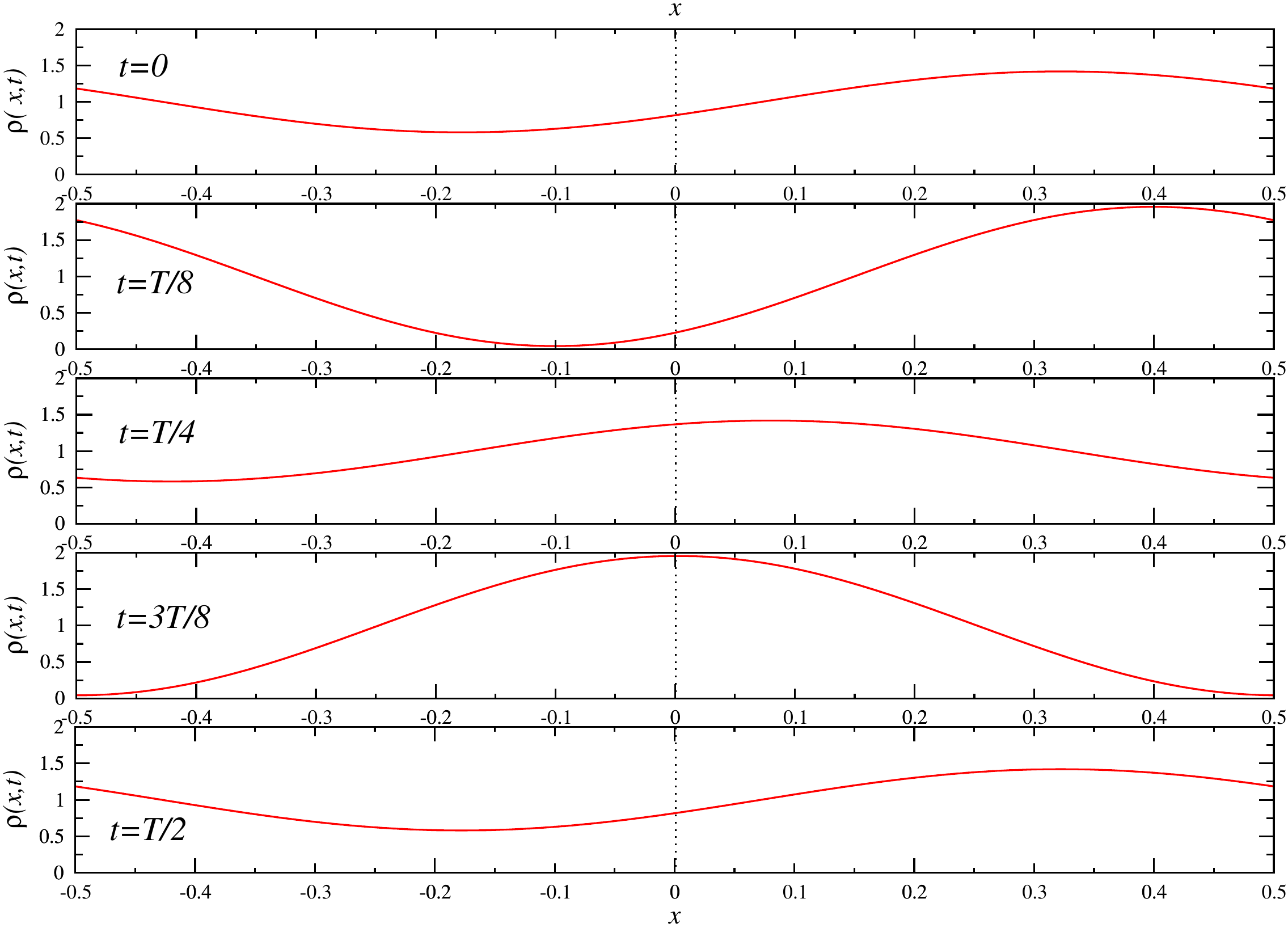,width=10cm}
\epsfig{file=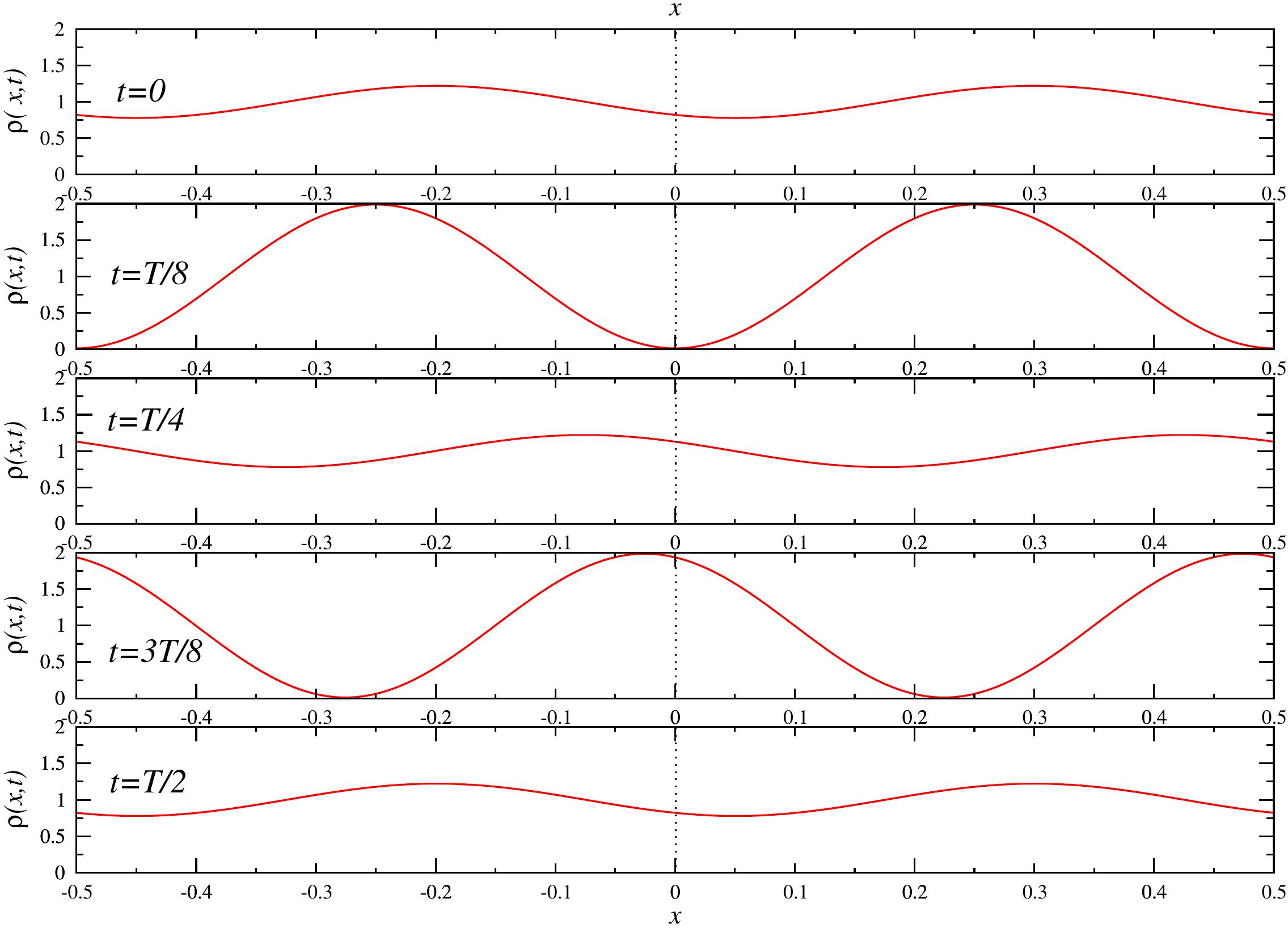,width=10cm}
\end{center}
\caption{\it Probability density $\rho(x,t)$ for a Majorana fermion confined to
a 1-d interval $x/L = [-\frac{1}{2},\frac{1}{2}]$ in the parity-violating case
($s_+ = s_-$) with $M L c = 1$ for various times $t$ in units of the period 
$T = 2 \pi/\omega$ of the wave function. Note that $\rho(x,t)$ is periodic in
time with period $T/2$. Since parity is now violated, 
$\rho(x,T/4) \neq \rho(-x,0)$. The state of lowest frequency is shown in the 
top panels, while the first excited state is shown at the bottom.}
\label{Pviolating}
\end{figure}

In the parity-violating case ($s_+ = s_-$), the probability density is given 
by a more complicated expression, which we don't display here explicitly. The 
corresponding probability density is illustrated in Fig.\ref{Pviolating}.

\section{Majorana Fermions in $(3+1)$ Dimensions}

In this section we extend our previous considerations from $(1+1)$ to $(3+1)$
dimensions. We again consider the Majorana equation and its symmetries as well 
as a mapping to a relativistic Schr\"odinger equation.

\subsection{The Majorana equation in $(3+1)$ dimensions}

We start out with the Dirac equation in $(3+1)$ dimensions
\begin{eqnarray}
&&i \partial_t \Psi(\vec x,t) (\vec \alpha \cdot \vec p c + \beta M c^2)
\Psi(\vec x,t), \quad \Psi(x,t) = \left(\begin{array}{c} 
\psi_1(x,t) \\ \psi_2(x,t) \\ \psi_3(x,t) \\ \psi_4(x,t)\end{array}\right), 
\nonumber \\
&&\vec \alpha = 
\left(\begin{array}{cc} 0 & \vec \sigma \\ \vec \sigma & 0 \end{array}\right), 
\quad \beta = \left(\begin{array}{cc} \1 & 0 \\ 0 & -\1 \end{array}\right). 
\end{eqnarray}
For the $\gamma$-matrices we choose the Dirac basis
\begin{equation}
\gamma^0 = \beta = \left(\begin{array}{cc} \1 & 0 \\ 0 & -\1 \end{array}\right),
\quad \gamma^i = \gamma^0 \alpha^i =
\left(\begin{array}{cc} 0 & \sigma^i \\ -\sigma^i & 0 \end{array}\right),
\end{equation}
where $\vec \sigma$ are the Pauli matrices and we use the space-time metric 
$g_{\mu\nu} = \mbox{diag}(1,-1,-1,-1)$. Next, we consider the Majorana basis
\begin{equation}
\widetilde \gamma^0 = 
\left(\begin{array}{cc} 0 & \sigma^2 \\ \sigma^2 & 0 \end{array}\right),
\widetilde \gamma^1 = 
\left(\begin{array}{cc} i \sigma^1 & 0 \\ 0 & i \sigma^1 \end{array}\right),
\widetilde \gamma^2 = 
\left(\begin{array}{cc} 0 & \sigma^2 \\ - \sigma^2 & 0 \end{array}\right),
\widetilde \gamma^3 = 
\left(\begin{array}{cc} i \sigma^3 & 0 \\ 0 & \sigma^3 \end{array}\right),
\end{equation}
in which the $\widetilde\gamma$ matrices again have purely imaginary entries. 
In this basis, the Majorana condition takes the simple form
\begin{equation}
\widetilde \Psi(\vec x,t) = \widetilde \Psi(\vec x,t)^*.
\end{equation}
The Dirac and the Majorana basis are now related by the unitary transformation
\begin{equation}
U = \frac{1}{2} 
\left(\begin{array}{cccc} 1 & -1 & -i & -i \\ 1 & 1 & i & -i \\
i & i & 1 & -1 \\ -i & i & 1 & 1 \end{array}\right), \quad
\gamma^\mu = U \widetilde\gamma^\mu U^\dagger, \quad 
\Psi(x,t) = U \widetilde\Psi(x,t).
\end{equation}
In the Dirac basis, the Majorana condition reads
\begin{eqnarray}
\label{Mcondition3d}
\Psi(\vec x,t)&=&U \widetilde\Psi(x,t) = U \widetilde\Psi(\vec x,t)^* =
U [U^\dagger \Psi(\vec x,t)]^* = U U^T \Psi(\vec x,t)^* \nonumber \\
&=&\frac{1}{4} \left(\begin{array}{cccc} 0 & 0 & 0 & -i \\ 0 & 0 & i & 0 \\
0 & i & 0 & 0 \\ -i & 0 & 0 & 0 \end{array}\right) \left(\begin{array}{c} 
\psi_1(\vec x,t)^* \\ \psi_2(\vec x,t)^* \\  \psi_3(\vec x,t)^* \\ 
\psi_4(\vec x,t)^*\end{array}\right) \ \Rightarrow \nonumber \\
\psi_3(\vec x,t)&=&i \psi_2(\vec x,t)^*, \quad 
\psi_4(\vec x,t) = -i \psi_1(\vec x,t)^* \Rightarrow \nonumber \\
&&\left(\begin{array}{c} \psi_3(\vec x,t) \\ \psi_4(\vec x,t) \end{array} 
\right) = - \sigma_2 \left(\begin{array}{c} \psi_1(\vec x,t)^* \\ 
\psi_2(\vec x,t)^* \end{array} \right).
\end{eqnarray}
Introducing the two-component Majorana spinor
\begin{equation}
\psi(\vec x,t) = \left(\begin{array}{c} \psi_1(\vec x,t) \\ \psi_2(\vec x,t)
\end{array}\right),
\end{equation} 
the 4-component Dirac equation reduces to the 2-component Majorana equation
\begin{eqnarray}
\label{Majorana3d}
&&i \partial_t 
\left(\begin{array}{c} \psi_1(\vec x,t) \\ \psi_2(\vec x,t) \\ 
i \psi_2(\vec x,t)^* \\ - i \psi_1(\vec x,t)^* \end{array}\right) =
\left(\vec \alpha \cdot \vec p c + \beta M c^2\right)  
\left(\begin{array}{c} \psi_1(\vec x,t) \\ \psi_2(\vec x,t) \\ 
i \psi_2(\vec x,t)^* \\ - i \psi_1(\vec x,t)^* \end{array}\right) \
\Rightarrow \nonumber \\
&&i \partial_t \psi(x,t) =  M c^2 \psi(x,t) - 
c \vec \sigma \cdot \vec p \sigma^2 \psi(x,t)^*.
\end{eqnarray}

\subsection{Conserved current}

Again, the Majorana equation inherits the conserved current of the Dirac 
equation, 
\begin{eqnarray}
j^\mu(\vec x,t)&=&c\overline{\Psi}(\vec x,t) \gamma^\mu \Psi(\vec x,t) =
(c \rho(\vec x,t),\vec j(\vec x,t)) \ \Rightarrow \nonumber \\
\rho(\vec x,t)&=&\overline{\Psi}(\vec x,t) \gamma^0 \Psi(\vec x,t) = 
\Psi(\vec x,t)^\dagger \Psi(\vec x,t), \nonumber \\
\vec j(\vec x,t)&=&c \overline{\Psi}(\vec x,t) \vec \gamma \Psi(\vec x,t) = 
c \Psi(\vec x,t)^\dagger \gamma^0 \vec \gamma \Psi(\vec x,t) = 
c \Psi(\vec x,t)^\dagger \vec \alpha \Psi(\vec x,t).
\end{eqnarray}
By imposing the Majorana condition eq.(\ref{Mcondition}), the charge and current
densities take the form
\begin{equation}
\label{Majoranacurrent3d}
\rho(\vec x,t) = 2 \psi(\vec x,t)^\dagger \psi(\vec x,t), \quad 
\vec j(\vec x,t) = 
- c \psi(\vec x,t)^\dagger \vec \sigma \sigma^2 \psi(\vec x,t)^*
- c \psi(\vec x,t)^T \sigma^2 \vec \sigma \psi(\vec x,t).
\end{equation}
By using the Majorana equation (\ref{Majorana1d}) it is again straightforward to
verify the continuity equation
\begin{equation}
\partial_t \rho(\vec x,t) + \vec \nabla \cdot \vec j(\vec x,t) = 0.
\end{equation}

\subsection{Lorentz invariance, parity, time-reversal, and charge conjugation}

Just as in $(1+1)$ dimensions, it is straightforward to show that the $(3+1)$-d
Majorana condition eq.(\ref{Mcondition3d}) is again Lorentz covariant. Let us
also consider the discrete symmetries P, T, and C for $(3+1)$-d Majorana 
fermions. For a Dirac fermion field $\Psi(\vec x,t)$, parity P corresponds to
$\gamma^0 \Psi(- \vec x,t)$. This transformation is again incompatible with the 
Majorana condition, but can be combined with a $U(1)$ phase multiplication by 
$i$, such that
\begin{eqnarray}
&&^P\Psi(\vec x,t) = i \gamma^0 \Psi(- \vec x,t) = 
i \left(\begin{array}{cc} \1 & 0 \\ 0 & - \1 \end{array}\right) \Psi(- \vec x,t)
\ \Rightarrow \nonumber \\
&&^P\psi_1(\vec x,t) = i \psi_1(- \vec x,t), \quad 
^P\psi_2(\vec x,t) = i \psi_2(- \vec x,t), \nonumber \\
&&^P\psi_3(\vec x,t) = - i \psi_3(- \vec x,t), \quad 
^P\psi_4(\vec x,t) = - i \psi_4(- \vec x,t).
\end{eqnarray}
Hence, for a $(3+1)$-d Majorana fermion parity takes the form
\begin{equation}
^P\psi(\vec x,t) = 
\left(\begin{array}{c} ^P\psi_1(\vec x,t) \\ ^P\psi_2(\vec x,t) \end{array}
\right) = 
\left(\begin{array}{c} i \ \psi_1(- \vec x,t) \\ i \psi_2(- \vec x,t) 
\end{array}\right) = i \psi(- \vec x,t).
\end{equation}
This transformation is consistent with the Majorana condition because
\begin{eqnarray}
&&^P \psi_3(\vec x,t) = - i \psi_3(- \vec x,t) = \psi_2(- \vec x,t)^* =
i \ ^P\psi_2(\vec x,t)^*, \nonumber \\
&&^P \psi_4(\vec x,t) = - i \psi_4(- \vec x,t) = - \psi_1(- \vec x,t)^* =
- i \ ^P\psi_1(\vec x,t)^*,
\end{eqnarray}
and it indeed leaves the Majorana equation invariant. This is straightforward
to show using
\begin{equation}
\sigma^2 (\vec \sigma \cdot \vec p)^* \sigma^2 = 
- \sigma^2 \vec \sigma^* \sigma^2 \cdot \vec p = \vec \sigma \cdot \vec p.
\end{equation}

For a Dirac fermion in $(3+1)$-d time-reversal takes the form
\begin{equation}
^T\Psi(\vec x,t) = \left(\begin{array}{cc} \sigma^2 & 0 \\ 0 & \sigma^2
\end{array}\right) \Psi(\vec x,- t)^*.
\end{equation}
For a Majorana spinor this implies
\begin{equation}
^T\psi(\vec x,t) = \sigma^2 \psi(\vec x,- t)^*.
\end{equation}
It is again straightforward to check that this transformation leaves the 
Majorana equation invariant. 

Finally, let us consider charge conjugation C, which for a $(3+1)$-d Dirac 
fermion takes the form
\begin{equation}
^C\Psi(\vec x,t) = \left(\begin{array}{cc} 0 & \sigma^2 \\ - \sigma^2 & 0 
\end{array}\right) 
\Psi(\vec x,t)^*.
\end{equation}
This implies that a Majorana fermion is indeed C-invariant
\begin{equation}
^C\psi(\vec x,t) = \psi(\vec x,t).
\end{equation}

\subsection{Relation of the $(3+1)$-d Majorana equation to a \\ relativistic 
Schr\"odinger equation}

Let us also consider the relation of the $(3+1)$-d Majorana equation to a 
relativistic Schr\"odinger equation. In this case we construct
\begin{equation}
\Phi(\vec x,t) = \psi(\vec x,t) - \frac{\vec \sigma \cdot \vec p c \sigma^2}
{\sqrt{(M c^2)^2 + p^2 c^2} + M c^2} \psi(\vec x,t)^*, \quad 
\vec p = - i \vec \nabla.
\end{equation}
It is straightforward to show that, just as in $(1+1)$-d, $\Phi(\vec x,t)$ obeys
the relativistic Schr\"odinger-type equation
\begin{equation}
i \partial_t \Phi(\vec x,t) = \sqrt{(M c^2)^2 + p^2 c^2} \ \Phi(\vec x,t).
\end{equation}
In this case, $\Phi(\vec x,t)$ is a 2-component spinor, which enjoys a global
$U(2)$ symmetry
\begin{equation}
\label{U2symmetry}
^\Omega\Phi(\vec x,t) = \Omega \Phi(\vec x,t), \quad \Omega \in U(2).
\end{equation}
This symmetry is not manifest in the Majorana equation. In fact, the $U(2)$
symmetry is like an internal ``flavor'' symmetry, while the two components of
the original Majorana spinor are related by space-time rotations. Again, we can 
invert the relation between $\psi$ and $\Phi$
\begin{equation}
\psi(\vec x,t) = \frac{1}{2 \sqrt{(M c^2)^2 + p^2 c^2}} \left[
\left(\sqrt{(M c^2)^2 + p^2 c^2} + M c^2\right) \Phi(\vec x,t) + 
\vec \sigma \cdot \vec p c \sigma^2 \Phi(\vec x,t)^*\right].
\end{equation}
The $U(2)$ symmetry of eq.(\ref{U2symmetry}) then turns into the nonlocal 
transformation
\begin{eqnarray}
^\Omega\psi(\vec x,t)&=&
\frac{1}{2 \sqrt{(M c^2)^2 + p^2 c^2}} 
\left[\left(\sqrt{(M c^2)^2 + p^2 c^2} + M c^2\right) \ ^\Omega\Phi(\vec x,t) 
\right. \nonumber \\
&+&\left. \vec \sigma \cdot \vec p c \sigma^2 \ ^\Omega\Phi(\vec x,t)^*\right] 
\nonumber \\
&=&\frac{1}{2 \sqrt{(M c^2)^2 + p^2 c^2}} 
\left[\left(\sqrt{(M c^2)^2 + p^2 c^2} + M c^2\right) \Omega \Phi(\vec x,t) 
\right. \nonumber \\
&+&\left. \vec \sigma \cdot \vec p c \sigma^2 \Omega^* \Phi(\vec x,t)^*\right] 
\end{eqnarray}

Just as in $(1+1)$-d, Lorentz invariance, which is manifest in the Majorana 
equation, is represented by a complicated nonlinear transformation of the 
Schr\"odinger-type wave function $\Phi$, which inherits its P and T symmetry 
properties from the Majorana ``wave function'' $\psi$
\begin{eqnarray}
^P \Phi(\vec x,t)&=& \ ^P\psi(\vec x,t) - 
\frac{\vec \sigma \cdot \vec p c \sigma^2}
{\sqrt{(M c^2)^2 + p^2 c^2} + M c^2} \psi(\vec x,t)^* \nonumber \\
&=&i \psi(- \vec x,t) + i \frac{\vec \sigma \cdot \vec p c \sigma^2}
{\sqrt{(M c^2)^2 + p^2 c^2} + M c^2} \psi(- \vec x,t)^* = i \Phi(- \vec x,t), 
\nonumber \\
^T \Phi(\vec x,t)&=& \ ^T\psi(\vec x,t) -
\frac{\vec \sigma \cdot \vec p c \sigma^2}
{\sqrt{(M c^2)^2 + p^2 c^2} + M c^2} \ ^T\psi(\vec x,t)^* \nonumber \\
&=&\sigma^2 \psi(\vec x,-t)^* + \frac{\vec \sigma \cdot \vec p c \sigma^2}
{\sqrt{(M c^2)^2 + p^2 c^2} + M c^2} \sigma^2 \psi(\vec x,-t) \nonumber \\
&=&\sigma^2 \Phi(x,-t)^*.
\end{eqnarray}

\section{Perfectly Reflecting Walls for $(3+1)$-d Majorana Fermions}

In this section we study the self-adjoint extension parameters that characterize
a perfectly reflecting boundary condition for $(3+1)$-d Majorana fermions.
Let us first consider Dirac fermions confined to a finite 3-d spatial domain 
$\Omega$ \cite{AlH12}. In order to investigate the Hermiticity of the 
Hamiltonian we consider
\begin{eqnarray}
\langle \chi|H|\Psi\rangle&=&\int_\Omega d^3x \ \chi(\vec x)^\dagger 
\left[\vec \alpha \cdot \vec p c + \beta M c^2\right] 
\Psi(\vec x) \nonumber \\
&=&\int_\Omega d^3x \ \chi(\vec x)^\dagger 
\left[\vec \alpha \cdot \left(- i \vec \nabla \right) c + \beta M c^2\right] 
\Psi(\vec x) \nonumber \\
&=&\int_\Omega d^3x \ \left\{\left[\vec \alpha \cdot 
\left(- i \vec \nabla \right) c + \beta M c^2\right]\chi(\vec x)\right\}^\dagger 
\Psi(\vec x) \nonumber \\
&-&i c \int_{\p \Omega} d\vec n \cdot \chi(\vec x)^\dagger \vec \alpha 
\Psi(\vec x) \nonumber \\
&=&\langle \Psi|H|\chi\rangle^* - 
i c \int_{\p \Omega} d\vec n \cdot \chi(\vec x)^\dagger \vec \alpha \Psi(\vec x),
\end{eqnarray}
which thus leads to the Hermiticity condition
\begin{equation}
\label{Dirachermiticity}
\chi(\vec x)^\dagger \vec n(\vec x) \cdot \vec \alpha \Psi(\vec x) = 0, 
\quad \vec x \in \p \Omega.
\end{equation}
Here $\vec n(\vec x)$ is the unit-vector normal to the boundary $\p \Omega$.
Next we introduce the self-adjoint extension condition
\begin{equation}
\label{Diracselfadjointness}
\left(\begin{array}{c} \Psi_3(\vec x) \\ \Psi_4(\vec x) \end{array}\right) = 
\lambda(\vec x) 
\left(\begin{array}{c} \Psi_1(\vec x) \\ \Psi_2(\vec x) \end{array}\right),
\quad \lambda(\vec x) \in GL(2,\C), \quad \vec x \in \p \Omega,
\end{equation}
which turns eq.(\ref{Dirachermiticity}) to
\begin{eqnarray}
&&\chi(\vec x)^\dagger 
\left(\begin{array}{cc} \0 & \vec n(\vec x) \cdot \vec \sigma \\ 
\vec n(\vec x) \cdot \vec \sigma & \0 \end{array}\right) \Psi(\vec x) = 
\nonumber \\
&&\left[\left(\chi_1(\vec x)^*,\chi_2(\vec x)^* \right) 
\vec n(\vec x) \cdot \vec \sigma \lambda(\vec x) + 
\left(\chi_3(\vec x)^*,\chi_4(\vec x)^* \right)
\vec n(\vec x) \cdot \vec \sigma \right]
\left(\begin{array}{c} \Psi_1(\vec x) \\ \Psi_2(\vec x) \end{array}\right) 
= 0 \ \Rightarrow \nonumber \\ 
&&\left(\begin{array}{c} \chi_3(\vec x) \\ \chi_4(\vec x) \end{array}\right) = 
- \vec n(\vec x) \cdot \vec \sigma \lambda(\vec x)^\dagger 
\vec n(\vec x) \cdot \vec \sigma
\left(\begin{array}{c} \chi_1(\vec x) \\ \chi_2(\vec x) \end{array}\right),
\end{eqnarray}
In order to guarantee self-adjointness of $H$, i.e.\ the equality of the domains
$D(H) = D(H^\dagger)$, we demand
\begin{equation}
\lambda(\vec x) = - \vec n(\vec x) \cdot \vec \sigma \lambda(\vec x)^\dagger 
\vec n(\vec x) \cdot \vec \sigma \ \Rightarrow \
\vec n(\vec x) \cdot \vec \sigma \lambda(\vec x) = -
\left[\vec n(\vec x) \cdot \vec \sigma \lambda(\vec x)\right]^\dagger.
\end{equation}
Hence, $\vec n(\vec x) \cdot \vec \sigma \lambda(\vec x)$ is anti-Hermitean. 
For Dirac fermions, there is thus a 4-parameter family of self-adjoint 
extensions that characterizes a perfectly reflecting wall. In the MIT bag
model \cite{Cho74,Cho74a,Has78}, the boundary condition was chosen as
$\lambda(\vec x) = i \vec n(\vec x) \cdot \vec \sigma$. This maintains spatial 
rotation invariance around the normal $\vec n(\vec x)$ on the boundary, but is 
not the most general choice.

Let us now impose the Majorana condition eq.(\ref{Mcondition3d}), which implies
\begin{eqnarray}
&&\lambda(\vec x) \left(\begin{array}{c} \Psi_1(\vec x,t) \\ \Psi_2(\vec x,t)
\end{array}\right) = \left(\begin{array}{c} \Psi_3(\vec x,t) \\ \Psi_4(\vec x,t)
\end{array}\right) =
- \sigma^2 \left(\begin{array}{c} \Psi_1(\vec x,t)^* \\ \Psi_2(\vec x,t)^*
\end{array}\right) \ \Rightarrow \nonumber \\
&&\lambda(\vec x)^* 
\left(\begin{array}{c} \Psi_1(\vec x,t)^* \\ \Psi_2(\vec x,t)^*
\end{array}\right) = 
\sigma^2 \left(\begin{array}{c} \Psi_1(\vec x,t) \\ \Psi_2(\vec x,t)
\end{array}\right) \ \Rightarrow \nonumber \\
&&\lambda(\vec x) \sigma^2 \lambda(\vec x)^*
\left(\begin{array}{c} \Psi_1(\vec x,t)^* \\ \Psi_2(\vec x,t)^*
\end{array}\right) = 
\lambda(\vec x) \left(\begin{array}{c} \Psi_1(\vec x,t)^* \\ \Psi_2(\vec x,t)^*
\end{array}\right) = 
- \sigma^2 \left(\begin{array}{c} \Psi_1(\vec x,t)^* \\ \Psi_2(\vec x,t)^*
\end{array}\right).
\end{eqnarray}
In order to be consistent with the Majorana condition eq.(\ref{Mcondition3d}),
the matrix $\lambda(\vec x)$ of self-adjoint extension parameters must hence 
obey
\begin{equation}
\lambda(\vec x) \sigma^2 \lambda(\vec x)^* = - \sigma^2.
\end{equation}
How does this constraint affect the original 4-parameter family of self-adjoint
extensions? In order to answer this question, let us perform a unitary 
transformation $V(\vec x) \in SU(2)$ that diagonalizes 
$\vec n(\vec x) \cdot \vec \sigma$, i.e.
\begin{eqnarray}
&&V(\vec x) \vec n(\vec x) \cdot \vec \sigma V(\vec x)^\dagger = \sigma^3, \quad
V(\vec x) \lambda(\vec x) V(\vec x)^\dagger = \lambda(\vec x)' \ \Rightarrow
\nonumber \\
&&\sigma^3 \lambda(\vec x)' = - \left[\sigma^3 \lambda(\vec x)'\right]^\dagger,
\quad \lambda(\vec x)' \sigma^2 \lambda(\vec x)^{'*} = - \sigma^2.
\end{eqnarray}
First of all, we make the ansatz
\begin{equation}
\lambda(\vec x)' = \lambda_0(\vec x)' + \vec \lambda(\vec x)' \cdot \vec \sigma,
\quad \lambda_0(\vec x)', \lambda_i(\vec x)' \in \C.
\end{equation}
The condition 
\begin{equation}
\sigma^3 \lambda(\vec x)' = - \left[\sigma^3 \lambda(\vec x)'\right]^\dagger,
\end{equation}
then implies
\begin{equation}
\lambda_0(\vec x)' = - \lambda_0(\vec x)'^*, \quad 
\lambda_1(\vec x)' = \lambda_1(\vec x)'^*, \quad 
\lambda_2(\vec x)' = \lambda_2(\vec x)'^*, \quad 
\lambda_3(\vec x)' = - \lambda_3(\vec x)'^*, 
\end{equation}
which indeed represents a 4-parameter family of self-adjoint extensions.
The additional relation
\begin{equation}
\label{condition}
\lambda(\vec x)' \sigma^2 \lambda(\vec x)^{'*} = - \sigma^2,
\end{equation}
can be satisfied in two different ways. First, we assume that
$\lambda_1(\vec x)' = \lambda_2(\vec x)' = 0$. In that case,
eq.(\ref{condition}) implies 
\begin{equation}
\lambda_0(\vec x)'^2 - \lambda_3(\vec x)'^2 = 1,
\end{equation}
which reduces the original 4-parameter family for Dirac fermions to a 
1-parameter family of self-adjoint extensions for Majorana fermions. 
Alternatively, we may assume that $\lambda_0(\vec x)' = \lambda_3(\vec x)' = 0$.
In that case, eq.(\ref{condition}) implies 
\begin{equation}
\lambda_1(\vec x)'^2 + \lambda_2(\vec x)'^2 = 1,
\end{equation}
which corresponds to another 1-parameter family of self-adjoint extensions.
Hence, we conclude that the boundary conditions for Majorana fermions confined 
to a finite 3-d spatial volume are characterized by two distinct 1-parameter
families of self-adjoint extensions.

\section{Conclusions}

Motivated by the edge modes of Kitaev wires or superconductors, as well as
by engineered quantum systems of ultracold atoms or trapped ions that
can be used as quantum simulators, we have investigated Majorana fermions 
confined to a 1-d interval or to a 3-d finite volume. This required an
understanding of the self-adjoint extension parameters that characterize the
most general perfectly reflecting boundary conditions. In contrast to $(1+1)$-d
Dirac fermions, whose hard wall boundary conditions are described by a 
continuous 1-parameter family of self-adjoint extension parameters, there are 
only two discrete types of wall boundary conditions for $(1+1)$-d Majorana 
fermions. In three spatial dimensions, on the other hand, the most general 
perfectly reflecting wall boundary condition for Dirac fermions is characterized
by a 4-parameter family of self-adjoint extension parameters, while the 
corresponding boundary condition for Majorana fermions is characterized by two 
different families of self-adjoint extensions, each with only a single 
parameter. Based on these results, one can derive the features of engineered
systems of Majorana fermions in a variety of confining spatial geometries, which
we did here explicitly for a 1-d interval. In addition, we have mapped the 
Majorana equation in one and three spatial dimensions to an equivalent nonlocal 
relativistic Schr\"odinger-type equation, whose quantum mechanical 
interpretation as a single-particle equation is not problematical.

\section*{Acknowledgments}

This publication was made possible by the NPRP grant \# NPRP 5 - 261-1-054 from 
the Qatar National Research Fund (a member of the Qatar Foundation). The 
statements made herein are solely the responsibility of the authors.

\end{document}